\documentclass[journal]{IEEEtran}

\ifCLASSINFOpdf
\else
   \usepackage[dvips]{graphicx}
\fi
\usepackage{url}

\usepackage{graphicx}

\usepackage{cite}
\usepackage{bm}
\usepackage{amsmath}
\usepackage{amssymb}
\usepackage{multirow}
\usepackage{enumitem}
\usepackage{booktabs}

\ifCLASSOPTIONcompsoc
 \usepackage[caption=false,font=normalsize,labelfont=sf,textfont=sf]{subfig}
\else
 \usepackage[caption=false,font=footnotesize]{subfig}
\fi
\begin{document}

% \title{Inaccurate Leaders-Assisted Joint Localization and Synchronization for 
% Dynamic Followers \\ Using Sequential TOA Measurements}
\title{Sequential TOA-Based Moving Target Localization in Multi-Agent Networks} 

\author{Qin Shi, Xiaowei Cui, Sihao Zhao and Mingquan Lu
\thanks{This paragraph of the first footnote will contain the date on which you submitted your paper for review.}
\thanks{Q. Shi, X. Cui and S. Zhao are with the Department of Electronic Engineering,
Tsinghua University, Beijing 100084, P. R. China (sqn175@gmail.com; cxw2005@tsinghua.edu.cn; 
zsh01@tsinghua.org.cn).}% <-this % stops a space
\thanks{M. Lu is with the Department of Electronic Engineering, Tsinghua University
and Beijing National Research Center for Information Science and Technology, Beijing, 100084, P. R. China
(lumq@tsinghua.edu.cn).}
}% <-this % stops a space

\markboth{ }
{Shell \MakeLowercase{\textit{et al.}}: Bare Demo of IEEEtran.cls for IEEE Journals}
\maketitle

\begin{abstract}
Localizing moving targets in unknown harsh environments has always been a severe challenge.
This letter investigates a novel localization system based on multi-agent networks, where multiple 
agents serve as mobile anchors broadcasting their time-space information to the targets.
We study how the moving target can localize itself using the sequential
time of arrival (TOA) of the one-way broadcast signals.   
An extended two-step weighted least squares (TSWLS) method is 
proposed to jointly estimate the position and velocity of the target
in the presence of agent information uncertainties.
We also address the large target clock offset (LTCO) problem for numerical stability.
Analytical results reveal that our method reaches the Cram\'er-Rao lower bound (CRLB) under small noises.
Numerical results show that the proposed method performs better than the existing algorithms.
\end{abstract}


\IEEEpeerreviewmaketitle

\section{Introduction}
\IEEEPARstart{M}{ulti-agent} networks have recently gained many attentions with their potential intelligent 
applications \cite{wang2019multi, shi2019blas}.
In the absence of a global navigation satellite system, networked agents can provide positioning services
by acting as moving anchors, such as temporary aerial and vehicular anchors in terrestrial emergency rescue. 
To be specific, the agents can periodically broadcast wireless signals and their time-space 
information \cite{shi2019blas, cano2019kalman, shi2019range}. 
The broadcast information has notable uncertainties since the networked agents can only self-determine 
their parameters in unknown environments.
The moving targets in communication range passively receive the signals and utilize the one-way time of arrival 
(TOA) measurements to localize themselves against the agents, and hence can theoretically scale to an arbitrary 
number \cite{shi2019blas}.

The TOA-based localization problem in the presence of anchor uncertainties is mainly studied 
for static scenarios in the area of wireless sensor network (WSN).
An intuitive maximum likelihood estimator (MLE) involving the parameters of both the static anchors 
and targets is proposed in \cite{angjelichinoski2014spear}. 
However, it cannot guarantee convergence without good initial guesses. 
Closed-form solutions, which do not require an initial guess, are reported in the literature
\cite{zheng2009joint,wang2015toa}.
In \cite{zheng2009joint}, a two-step generalized total least squares (TSGTLS) method is proposed 
to estimate the position of a static target using all two-way TOA measurements between the target and 
anchors.
In \cite{wang2015toa}, the two-step weighted least squares (TSWLS) localization method is 
derived to estimate the position of the target.
The method first linearizes the observation equations and then solves the problem in the sense of WLS 
by incorporating the agent uncertainties into the weighting matrix.
However, the above methods are still all discussed from the static WSN point of
view, which do not apply to moving targets.
For moving targets, the state parameters of practical interest consist not only the position,
but also the velocity of the target.

In practical implementation, channel access technology is used to permit collision-free signal broadcasting
from different anchors. 
For easy deployment using the off-the-shelf products, such as the recent ultra-wideband (UWB) chips, a time division 
multiple access (TDMA) scheme is commonly used \cite{shi2019blas,tiemann2019atlas}. 
Different from the two-way or concurrent TOA measurements in \cite{zheng2009joint,wang2015toa},
TDMA produces sequential one-way TOA measurements,
which is energy efficient for moving targets and capable of multiple targets localization.
However, the minor time difference between sequential measurements will degrade localization performance 
if not considered.
Fortunately, the minor time difference renders the target speed observable.
Therefore, further consideration of sequential TOA measurements is needed for moving target localization.

In this letter, we extend the TSWLS algorithm \cite{wang2015toa} to the more general case with moving targets as well 
as sequential TOA measurements.
In our localization system, the moving targets only passively receive signals from the multi-agent 
network that broadcasts information based on TDMA scheduling.
The proposed method only utilizes one-way TOA measurements to jointly estimate the 
position and velocity of the target, and thus is energy efficient and distributed.
To improve performance, we additionally utilize an optimization process to refine the estimates.
Furthermore, we introduce the new concept of large target clock offset (LTCO) scenarios concerning 
the unsynchronized targets that first enter or re-enter the multi-agent networks.
The scenarios are similar to the large equal radius (LER) conditions in TOA-based localization 
\cite{romero2008large, romero2011evaluation}, 
since they both lead to large approximately equal pseudoranges and hence induce ill-conditioned matrix.
We specially consider the LTCO scenarios by introducing a QR factorization to retain numerical stability.

\section{System Model and Problem Formulation}
The agents generally have onboard navigation systems, such as visual navigation systems and 
inertial navigation systems, and can exchange their measurements using wireless communication 
to collaboratively estimate their time-space information.
By periodically broadcasting their determined time-space information and corresponding uncertainties, 
the agents can serve 
as mobile anchors to assist low-cost moving target localization.
The passive targets can then obtain the TOA measurements of the broadcast signals and the agents' information 
to localize themselves.
Consider a multi-agent network $\mathcal M=\{1,2,\cdots,M\}$ and a target to be localized 
in a two-dimensional plane. 
To be specific, an agent $m \in \mathcal M$ will broadcast its real-time determined time-space information
$\hat {\boldsymbol\beta}_m = [\hat{\mathbf p}_m^T, \hat T_m]^T$ at its allocated broadcast time 
$t_{tx}^{m\rightarrow}$, where $\hat{\mathbf p}_m$ is its position and $\hat T_m$ is the  
clock offset with respect to the time reference agent $r$.
The TOA measurements can then be obtained by subtracting the arrival time recorded at the 
target from the broadcast time recorded at the agents. 
We aim to localize the target using the agent broadcast information and 
the sequential TOA measurements in one TDMA frame.
The target parameter vector at the start time of a TDMA frame is to be estimated and given as
\begin{equation}
   \mathbf x = [\mathbf p^T, \mathbf v^T, T, \omega]^T \in \mathbb R^6,
\end{equation}
where $\mathbf p=[x,y]^T$ is the target position, $\mathbf v=[v_x, v_y]^T$ is the velocity, $T$ and $\omega$ are the 
clock offset and clock skew with respect to agent $r$, respectively. 
The clock offset is the phase difference and the clock skew is the frequency difference 
of the internal clocks.

Under the first-order model assumption of the target motion and clock dynamic,
and the fact that the worst clock skew magnitude is up to $20$ parts per million (ppm) \cite{neirynck2016alternative}, 
the TOA measurement from agent $m\in \mathcal M$ can be modeled as \cite{shi2019blas}
\begin{equation} \label{eq:nonlinear toa}
   \tilde \tau_m = \tau_m + \Delta \tau_m, 
\end{equation}
\begin{equation} \label{eq:taum}
\tau_m = \Vert \mathbf p + \mathbf v t_m - \mathbf p_m\Vert 
+ T + \omega t_m - T_m,
\end{equation}
where $\Delta \tau_m$ denotes the independent measurement noise \cite{elson2002fine}, 
$\Vert \cdot \Vert$ denotes the $l_2$ norm, and
$t_m = t_{tx}^{m\rightarrow} - t_{tx}^{1\rightarrow}$ denotes the known slot time difference and is determined by 
TDMA designers, where $t_{tx}^{1\rightarrow}$ is the start time of a TDMA frame. 
Time-related variables are represented in its range-equivalent form by multiplying the known signal 
propagation speed. 

For general discussion, the TOA measurements are collected into 
$\tilde {\boldsymbol\tau} = [\tilde \tau_1, \tilde \tau_2, \cdots, \tilde\tau_M]^T$, which forms 
\begin{equation}
   \tilde {\boldsymbol\tau} = \boldsymbol \tau + 
      \Delta \boldsymbol \tau, 
\end{equation}
where $\Delta \boldsymbol \tau$ is modeled as zero-mean Gaussian random vector with diagonal covariance matrix 
$\mathbf C_{\boldsymbol \tau}$. 
The agent nominal parameters are grouped into 
$\boldsymbol \beta = [\boldsymbol \beta_1^T, \boldsymbol \beta_2^T, \cdots, \boldsymbol \beta_M^T]^T$,
where $ \boldsymbol \beta_m = [\mathbf p_m^T, T_m]^T$.
They are unknown and corrupted with errors.
We denote the error vector as
\begin{equation}
   \Delta \boldsymbol \beta = \hat{\boldsymbol \beta} - \boldsymbol \beta = 
   [\Delta \boldsymbol \beta_1^T, \Delta \boldsymbol \beta_2^T,  \cdots, \Delta \boldsymbol \beta_M^T]^T,
\end{equation}
where $\Delta \boldsymbol \beta_m = [\Delta \mathbf p_m^T, \Delta T_m]^T$ with
\begin{equation} \label{eq:delta param}
   \Delta \mathbf p_m = \hat{\mathbf p}_m - \mathbf p_m, \hspace{1em}
   \Delta T_m = \hat T_m - T_m.
\end{equation}
We assume that $\Delta \boldsymbol \beta$ is modeled as zero-mean Gaussian random vector with covariance matrix 
$\mathbf C_{\boldsymbol \beta}$. 
We note that $\mathbf C_{\boldsymbol \beta}$ is also broadcast to the target.
We also assume that $\Delta \boldsymbol \beta$ is independent of the measurement 
noises $\Delta \boldsymbol \tau$.
Let $\boldsymbol \eta = [\mathbf x^T, \boldsymbol \beta^T]^T \in \mathbb R^{6+3M}$,
the joint probability density function (PDF) of the observed TOAs, agent 
positions and clock offsets parameterized on unknown $\boldsymbol \eta$ is then given as
\begin{multline} \label{eq:PDF}
   p(\tilde{\boldsymbol \tau}, \hat {\boldsymbol \beta}; \boldsymbol \eta) = 
   \frac{1}{\vert \mathbf C_{\boldsymbol \tau} \vert^{1/2}}
   \exp \Big( - \frac {1}{2}(\tilde {\boldsymbol \tau} - \boldsymbol \tau)^T\mathbf C_{\boldsymbol \tau}^{-1}
   (\tilde {\boldsymbol \tau} - \boldsymbol \tau)  \Big)\\ 
   \cdot\frac{1}{(2\pi)^{2M} \vert \mathbf C_{\boldsymbol \beta} \vert^{1/2}}
   \exp \Big(- \frac {1}{2}(\hat{\boldsymbol \beta} - \boldsymbol \beta)^T\mathbf C_{\boldsymbol \beta}^{-1}
   (\hat{\boldsymbol \beta} - \boldsymbol \beta) \Big).
\end{multline}

We note that MLE involves the maximization with respect to parameters of both the 
unknown target and agents. 
The maximization can be a very high-dimensional problem if the number of agents is large. 
Furthermore, MLE is computationally expensive and needs proper initial guesses. 
Instead, we extend the TSWLS method and propose a low complexity method that requires no initial guesses.

\section{Proposed Method}
The first step of the proposed method reparametrizes the localization problem in a higher-dimensional space
to linearize the equations and solves the linear equations by a WLS estimator. 
The second step introduces a nonlinear optimization process to refine the target parameters by exploiting their 
relationship to the higher-dimensional solution.
The details are described as follows:

\textit{1) Step I:}
The nominal parameters in (\ref{eq:nonlinear toa}) are replaced using (\ref{eq:delta param}).
Defining $\hat \alpha_m \triangleq \tilde \tau_m + \hat T_m$,
re-arranging (\ref{eq:nonlinear toa}) and squaring both sides as in \cite{wang2015toa}, 
the linear equation can be derived as:
\begin{multline} \label{eq:pseudo-linear equation}
   2 \hat {\mathbf p}_m^T \mathbf p + 2 t_m \hat{\mathbf p}_m^T\mathbf v
   -2 \hat \alpha_m T -2  t_m \hat \alpha_m \omega 
   + (T^2 - \Vert \mathbf p\Vert^2) +  t_m^2 \cdot \\ (\omega^2 - \Vert \mathbf v \Vert^2)
   + 2 t_m (T \omega - \mathbf p^T\mathbf v) 
   = \Vert \hat {\mathbf p}_m\Vert^2 - \hat \alpha_m^2 + e_m ,
\end{multline}
where
\begin{equation} \label{eq:err}
   e_m = \mathbf b_m \Delta \boldsymbol \beta_m + d_m \Delta \tau_m + \mathcal O,
\end{equation}
where $\mathbf b_m = [2 (\mathbf p + \mathbf v t_m   - \hat{\mathbf p}_m)^T, d_m]$,
$d_m = - 2 (T + \omega t_m - \hat \alpha_m)$ and 
$\mathcal O$ represents the second-order error terms that will be ignored. 
Note that $e_m$ is associated with the agent nominal parameters and their uncertainties. 
We address the uncertainties issue by treating agent uncertainties as fine-tunings in $e_m$. 

Define the reparametrized vector 
$\boldsymbol \theta = [\mathbf x^T, \theta_1, \theta_2, \theta_3]^T \in \mathbb R^9$, where  
$\theta_1 = T^2 - \Vert \mathbf p\Vert^2$, $\theta_2 = \omega^2 - \Vert \mathbf v \Vert^2$ and
$\theta_3 = T \omega - \mathbf p^T\mathbf v$. 
The matrix form of (\ref{eq:pseudo-linear equation}) is given as
\begin{equation} \label{eq:pseudo-linear matrix}
   \mathbf A \boldsymbol \theta = \mathbf y + \mathbf e,
\end{equation}
where $\mathbf A = [\mathbf a_1^T, \mathbf a_2^T, \cdots$, $\mathbf a_M^T]^T,
\mathbf y = [y_1, y_2, \cdots, y_M]^T$
and $\mathbf e = \mathbf B \Delta \boldsymbol \beta + \mathbf D \Delta \boldsymbol \tau$
with
\begin{align*}
   \mathbf a_m &= [2 \hat {\mathbf p}_m^T, 2 t_m \hat{\mathbf p}_m^T, -2 \hat \alpha_m, -2 t_m \hat \alpha_m, 1, t_m^2, 2 t_m], \\
   y_m &= \Vert \hat {\mathbf p}_m\Vert^2 - \hat \alpha_m^2, \\
   \mathbf B &= \mathrm{diag}([\mathbf b_1, \mathbf b_2, \cdots, \mathbf b_M]), \\
   \mathbf D &= \mathrm{diag}([d_1, d_2, \cdots, d_M]).
\end{align*}
The covariance matrix of the error vector $\mathbf e$ is derived as 
\begin{equation} \label{eq:Ce}
   \mathbf C_e = \mathbf B \mathbf C_{\boldsymbol \beta} \mathbf B^T + 
                     \mathbf D \mathbf C_{\boldsymbol \tau} \mathbf D^T.
\end{equation}

Applying the whitening transformation to $\mathbf e$ by multiplying $\mathbf W = \mathbf C_e^{-1/2}$ to both 
sides of (\ref{eq:pseudo-linear matrix}), we have
\begin{equation} \label{eq:whitened}
   \mathbf W \mathbf A \boldsymbol \theta = \mathbf W \mathbf y + \mathbf W \mathbf e.
\end{equation}
The WLS solution to (\ref{eq:pseudo-linear matrix}) is then given as 
\begin{equation} \label{eq:wls solution}
   \hat {\boldsymbol \theta}_{\mathrm {WLS}} = (\mathbf A^T \mathbf C_e^{-1} \mathbf A)^{-1} 
                  \mathbf A^T \mathbf C_e^{-1} \mathbf y.
\end{equation}
The covariance matrix of the estimate is derived as \cite{kay1993fundamentals}
\begin{equation} \label{eq:C wls}
   \mathbf C_{\mathrm {WLS}}=(\mathbf A^T \mathbf C_e^{-1} \mathbf A)^{-1}.
\end{equation}

However, we notice that the moving targets typically have low-cost clocks due to their size and cost constraints.
Therefore, the clock offset of the target can be relatively large, such as exceeding one millisecond, if not well synchronized.
This makes pairwise $\hat \alpha_m$ approximately equal and the coefficient matrix $\mathbf A$ have a very large condition 
number and, consequently, yields a poor estimate.
We define these situations as LTCO scenarios.
To the best of the authors' knowledge, the LTCO scenarios have not been reported in previous closed-form solutions for 
TOA-based localization problems.
LTCO scenarios may appear for a long-running target if its clock is not promptly adjusted and synchronized. 
Moreover, the LTCO scenario occurs if an unsynchronized target first enters the multi-agent network. 
In this case, the target will have a large clock offset with respect to the agents.

We address LTCO scenarios by applying a reduced QR factorization to $\mathbf W \mathbf A$ in 
(\ref{eq:whitened}), with column pivoting, to get 
\begin{equation} \label{eq:qr}
   \mathbf Q \mathbf R \mathbf P^T \boldsymbol \theta = \mathbf W \mathbf y + \mathbf W \mathbf e,
\end{equation}
where $\mathbf Q$ is an $M\times 9$ orthogonal matrix, $\mathbf R$ is an $9 \times 9$ upper triangular matrix and 
$\mathbf P$ is an $9 \times 9$ permutation matrix \cite{hough1997complete}.
To derive the WLS solution $\hat {\boldsymbol \theta}_{\mathrm WLS}$, we solve the following equation by back substitution:
\begin{equation}
   \mathbf R \mathbf P^T \boldsymbol \theta = \mathbf Q^T \mathbf W \mathbf y.
\end{equation}
The covariance matrix of the estimate is derived as:
\begin{equation}
   \mathbf C_{\mathrm WLS} = (\mathbf P (\mathbf R^T \mathbf R) \mathbf P^T)^{-1}.
\end{equation}
This dramatically enhances numerical stability properties.
Our analysis is presented in terms of (\ref{eq:whitened}), but our practical algorithm implementation is based on (\ref{eq:qr}).

\textit{2) Step II:}
This step retracts the target parameters $\mathbf x$ from the WLS solution $\hat{\boldsymbol \theta}_{\mathrm {WLS}}$ 
by exploiting their nonlinear relationship. The relationship is directly given as follows:
\begin{align} \label{eq:optimization}
   \hat {\boldsymbol \theta}_{\mathrm {WLS}} 
   & = 
   \begin{bmatrix}
      \mathbf x \\
      T^2 - x^2 - y^2 \\
      \omega^2 - v_x^2 - v_y^2 \\
      T \omega - x v_x - y v_y
   \end{bmatrix} + \mathbf e_{\mathrm {WLS}} \\
   & = \mathbf f(\mathbf x) + \mathbf e_{\mathrm {WLS}},
\end{align}
where $\mathbf e_{\mathrm {WLS}}$ is the error vector with covariance matrix $\mathbf C_{\mathrm {WLS}}$.
The target parameters $\mathbf x$ can be estimated by solving the following optimization problem:
\begin{equation}
   \hat {\mathbf x} = \underset{\mathbf x}{\mathrm{arg\,min}}\, \left\{
   \Vert\hat {\boldsymbol \theta}_{\mathrm {WLS}} - \mathbf f(\mathbf x)\Vert^2_{\mathbf C_{\mathrm {WLS}}^{-1}} \right\},
\end{equation}
where $\Vert \mathbf v \Vert_{\mathbf W}=\Vert \mathbf W \mathbf v \Vert$ is the weighted $l_2$ norm. 
This can be solved using the iterative weighted least squares (IWLS) based on the Gauss-Newton numerical algorithm.
To be specific, the initial guess of $\mathbf x$ is obtained from the truncated $\hat {\boldsymbol \theta}_{\mathrm {WLS}}$.
The increment $\Delta \mathbf x = \mathbf x - \hat {\mathbf x}$ in each iteration is estimated 
as 
\begin{equation} \label{eq:delta x}
   \hat {\Delta \mathbf x} = (\mathbf J^T\mathbf C_{\mathrm {WLS}}^{-1} \mathbf J)^{-1}\mathbf J^T\mathbf C_{\mathrm {WLS}}^{-1} \mathbf r,
\end{equation}
where 
\begin{equation} \label{eq:delta x info}
   \begin{split} 
   \mathbf r &= \hat {\boldsymbol \theta}_{\mathrm {WLS}} - \mathbf f(\hat{\mathbf x}), \hspace{2em}
   \mathbf J = \begin{bmatrix} \mathbf I_6 \\ \mathbf J_1 \end{bmatrix}, \\
   \mathbf J_1 &= 
      \begin{bmatrix}
      -2 \hat x & -2 \hat y & 0 & 0 &2\hat T & 0 \\
      0 & 0 & -2\hat v_x & -2 \hat v_y & 0 & 2\hat \omega \\
      -\hat v_x & -\hat v_y & -\hat x & -\hat y & \hat \omega & \hat T
      \end{bmatrix},
   \end{split}
\end{equation}
where $\mathbf I_n$ denotes an $n\times n$ identity matrix in this letter. 
The results can then be updated as
\begin{equation}
   \hat {\mathbf x} \leftarrow \hat{\Delta \mathbf x} + \hat {\mathbf x}.
\end{equation}

\textit{Remark 1:} $\mathbf C_e$ depends on target nominal parameters, which are not available. 
In practical implementation, we first set $\mathbf C_e$ to be $\mathbf I_M$ to obtain an initial 
estimate $\boldsymbol \theta_{\mathrm{LS}}$. 
We then use $\mathbf x_{\mathrm{LS}}$ from truncated $\boldsymbol \theta_{\mathrm{LS}}$ to evaluate $\mathbf C_e$ and derive the WLS solution.

\textit{Remark 2:} The stopping criteria in IWLS is set as 
$\Vert \hat{\Delta \mathbf x}(1:2)\Vert^2 \leq \sum_{m=1}^M \mathrm{tr}(\mathbf C_{\mathbf p_m}) / M$,
where $\mathrm{tr}(\mathbf C_{\mathbf p_m})$ is the trace of the agent position covariance matrix
that can be extracted from $\mathbf C_{\boldsymbol \beta}$.
The maximum number of iterations is set to be $5$.

\textit{Remark 3:} If we ignore the speed and clock skew parameters, do not consider the LTCO scenarios, and 
set IWLS iterations to be one, our method degrades into the original TSWLS algorithm in \cite{wang2015toa}.

\section{Performance Analysis}
In this section, we analyze the CRLB of the localization problem and the mean square error (MSE) of 
our proposed estimator to theoretically prove the effectiveness of our method.

The natural logarithm of the joint PDF 
$p(\tilde{\boldsymbol \tau}, \hat {\boldsymbol \beta}; \boldsymbol \eta)$ is given as
\begin{multline}
   \ln p(\tilde{\boldsymbol \tau}, \hat {\boldsymbol \beta}; \boldsymbol \eta) =
    c - \frac {1}{2}(\tilde {\boldsymbol \tau} - \boldsymbol \tau)^T\mathbf C_{\boldsymbol \tau}^{-1}
    (\tilde {\boldsymbol \tau} - \boldsymbol \tau) \\ 
    - \frac {1}{2}(\hat{\boldsymbol \beta} - \boldsymbol \beta)^T\mathbf C_{\boldsymbol \beta}^{-1}
    (\hat{\boldsymbol \beta} - \boldsymbol \beta),
\end{multline}
where $c$ is a constant independent of $\boldsymbol \eta$. 
The CRLB of $\boldsymbol \eta$ is derived as \cite{kay1993fundamentals}
\begin{equation}
   \mathrm{CRLB}(\boldsymbol \eta) = -\left(\mathbb E \left[ \frac{\partial ^2 \ln p}
                        {\partial \boldsymbol \eta \partial \boldsymbol \eta^T} \right] \right)^{-1} 
   = 
   \begin{bmatrix}
      \mathbf R_1 & \mathbf R_2 \\ \mathbf R_2^T & \mathbf R_3
   \end{bmatrix}^{-1},
\end{equation}
where 
\begin{equation} \label{eq:R123}
\begin{split}
\mathbf R_1 &= -\mathbb E\left[\frac{\partial^2 \ln p}{\partial \mathbf x \partial \mathbf x^T} \right] = 
\left(\nabla_{\mathbf x} \boldsymbol \tau \right)^T\mathbf C_{\boldsymbol \tau}^{-1} 
\left(\nabla_{\mathbf x} \boldsymbol \tau \right),\\
\mathbf R_2 &= -\mathbb E\left[\frac{\partial^2 \ln p}{\partial \mathbf x \partial {\boldsymbol \beta}^T} \right] =  
\left(\nabla_{\mathbf x} \boldsymbol \tau \right)^T \mathbf C_{\boldsymbol \tau}^{-1} 
\left(\nabla_{\boldsymbol \beta} \boldsymbol \tau \right) ,\\
\mathbf R_3 &= -\mathbb E\left[\frac{\partial^2 \ln p}{\partial \boldsymbol \beta \partial {\boldsymbol \beta}^T} \right] =  
\left(\nabla_{\boldsymbol \beta} \boldsymbol \tau \right)^T \mathbf C_{\boldsymbol \tau}^{-1} 
\left(\nabla_{\boldsymbol \beta} \boldsymbol \tau \right) + \mathbf C_{\boldsymbol \beta}^{-1},   
\end{split}
\end{equation}
where the operator $\nabla_{\mathbf a} $ yields the row vector partial derivative of function $\mathbf f(\mathbf a)$ as:
\begin{equation}
   \nabla_{\mathbf a} \mathbf f = \frac{\partial \mathbf f}{\partial \mathbf a^T}.
\end{equation}
The partial derivatives above are derived using (\ref{eq:taum})
\begin{align}
   \nabla_{\mathbf x} \boldsymbol \tau &= 
      \left[
         \nabla_{\mathbf x} \tau_1^T,
         \nabla_{\mathbf x} \tau_2^T, \cdots, 
         \nabla_{\mathbf x} \tau_M^T
      \right]^T, \\
   \nabla_{\boldsymbol \beta} \boldsymbol \tau &= 
      \mathrm{diag(\left[
         \nabla_{\boldsymbol \beta_1} \tau_1, 
         \nabla_{\boldsymbol \beta_2} \tau_2, \cdots, 
         \nabla_{\boldsymbol \beta_M} \tau_M 
      \right])},
\end{align}
where 
\begin{equation}
   \nabla_{\mathbf x} \tau_m = 
   \left[\boldsymbol \rho_m^T,t_m \boldsymbol \rho_m^T, 1, t_m\right],
   \nabla_{\boldsymbol \beta_m} \tau_m =
   \left[-\boldsymbol \rho_m^T, -1 \right], 
\end{equation}
with $\boldsymbol \rho_m = (\mathbf p + \mathbf v t_m - \mathbf p_m) / \Vert \mathbf p + \mathbf v t_m - \mathbf p_m \Vert$.
The CRLB for $\mathbf x$ is then derived as in \cite{zheng2009joint}
\begin{align} \label{eq:CRLB}
   \mathrm{CRLB}(\mathbf x) &= \left(\mathbf R_1-\mathbf R_2\mathbf R_3^{-1}\mathbf R_2^T\right)^{-1}.
\end{align}

We then derive the analytical MSE of our proposed method.
From the second step, the estimate bias is 
\begin{equation} \label{eq:est bias}
   \mathrm{b}(\mathbf x) = \mathbb E(\hat{\mathbf x}) - \mathbf x
   = (\mathbf J^T\mathbf C_{\mathrm {WLS}}^{-1} \mathbf J)^{-1}\mathbf J^T
   \mathbf C_{\mathrm {WLS}}^{-1} \Delta \boldsymbol \theta_{\mathrm{WLS}},
\end{equation}
where $\Delta \boldsymbol \theta_{\mathrm{WLS}}$ is the estimate bias in the first step,
$\Delta \boldsymbol \theta_{\mathrm{WLS}} = (\mathbf A^T \mathbf C_e^{-1} \mathbf A)^{-1}\mathbf A^T \mathbf C_e^{-1}\mathbf e$.
Substituting (\ref{eq:C wls}) and $\Delta \boldsymbol \theta_{\mathrm{WLS}}$ into (\ref{eq:est bias}), we can obtain 
\begin{equation}
   \mathrm{b}(\mathbf x)
   = \left((\mathbf A \mathbf J)^T \mathbf C_e^{-1} \mathbf A \mathbf J\right)^{-1}
   (\mathbf A \mathbf J)^T \mathbf C_e^{-1} \mathbf e.
\end{equation}
Under the assumption that $\mathbf e$ is small enough,
\begin{equation}
   \mathrm{b}(\mathbf x) \simeq \mathbf 0.
\end{equation}
The covariance of the estimate of $\mathbf x$ is then derived as 
\begin{equation}
   \mathrm{cov}(\mathbf x) = \left((\mathbf A \mathbf J)^T \mathbf C_e^{-1} \mathbf A \mathbf J\right)^{-1}.
\end{equation}
From (\ref{eq:Ce}) and using matrix inversion lemma \cite{kay1993fundamentals}, we have 
\begin{multline} \label{eq:Cx inv}
   \mathrm{cov}(\mathbf x) =\Big( \mathbf G_2^T \mathbf C_{\boldsymbol \tau}^{-1} \mathbf G_2 -
   \left(\mathbf G_2^T\mathbf C_{\boldsymbol \tau}^{-1}\mathbf G_1\right)  \\
   \cdot(\mathbf G_1^T\mathbf C_{\boldsymbol \tau}^{-1}\mathbf G + \mathbf C_{\boldsymbol \beta}^{-1})^{-1} 
   \left(\mathbf G_2^T\mathbf C_{\boldsymbol \tau}^{-1}\mathbf G_1\right)^T \Big)^{-1},
\end{multline}
where $\mathbf G_2 = \mathbf D^{-1}\mathbf A \mathbf J$ and $\mathbf G_1 = \mathbf D^{-1} \mathbf B$.
Comparing (\ref{eq:Cx inv}) with (\ref{eq:R123}) and (\ref{eq:CRLB}), using some manipulations,
it can be found that 
\begin{equation}
   \mathrm{cov}(\mathbf x)  \approx \mathrm{CRLB}(\mathbf x),
\end{equation}
when the measurement noises and agent errors are small and hence the elements in $\mathbf A, \mathbf B$ and $\mathbf D$ 
are with small perturbations. 
To conclude, our method is approximately unbiased and reaches CRLB 
under small noises.

\section{Numerical Results}
This section numerically evaluates the performance of our proposed method.
The simulation scenario is given in \mbox{Fig. \ref{fig:qin1}}, where a target and a multi-agent network composed of $M=10$ agents
are moving in a two-dimensional space.
Without loss of generality, agent $1$ is chosen as the time reference agent.
We set 10 time slots for TDMA scheduling with equal slot interval \mbox{$0.05$ s}, and the first slot 
starts at time \mbox{$0$ s}.
The agent clock offsets are generated by continuous uniform distribution, 
\mbox{$T_m \sim \mathcal U(-10, 10)$ ns}.
The clock skews of the target and agents are from $-20$ to \mbox{$20$ ppm}.
The target and agents move with a common velocity of \mbox{$(-5, 0)$ m/s}.
For simplicity, the measurement noise $\Delta \tau_m$ and the errors in agent position and clock offset
are assumed as independent zero-mean Gaussian random variables. 
Considering for example the UWB technology with fine temporal resolution, 
the covariance of $\Delta \tau_m$ is set as \mbox{$\sigma_{\tau}^2 = -30$ dB} \cite{xu2008delay}.
The covariance matrix of agent uncertainties is 
$\mathbf C_{\boldsymbol \tau} = \mathrm{diag}(\mathbf C_1, \mathbf C_2, \cdots, \mathbf C_M)$ where 
$\mathbf C_m = \sigma_m^2 \mathbf I_3$, and $\sigma_m^2$ follows a uniform distribution
\mbox{$\sigma_m^2 \sim \mathcal U(\sigma_s^2 - 5, \sigma_s^2 + 5)$ dB}. 
The MSEs of the results are evaluated by averaging over $N=3000$ independent runs.
For comparison, the TSGTLS in \cite{zheng2009joint} and the TSWLS in \cite{wang2015toa} are also simulated. 
Also included is the MLE using Gauss-Newton implementation
that ignores the agent uncertainties \cite{shi2019blas}. 
The initial guess for MLE is obtained by adding a small zero-mean Gaussian noise to the nominal parameters.
However, we note that the above three methods are all originally designed for a stationary target and are not 
applicable to our simulation scenarios.
For a fair comparison, we extensively implement these methods by considering the target speed and clock skew parameters 
using the sequential TOA measurements.
The implementations are based on our system model.
Three simulation schemes are devised as follows:

\begin{figure}
   \centering  
   \includegraphics[width=0.8\linewidth]{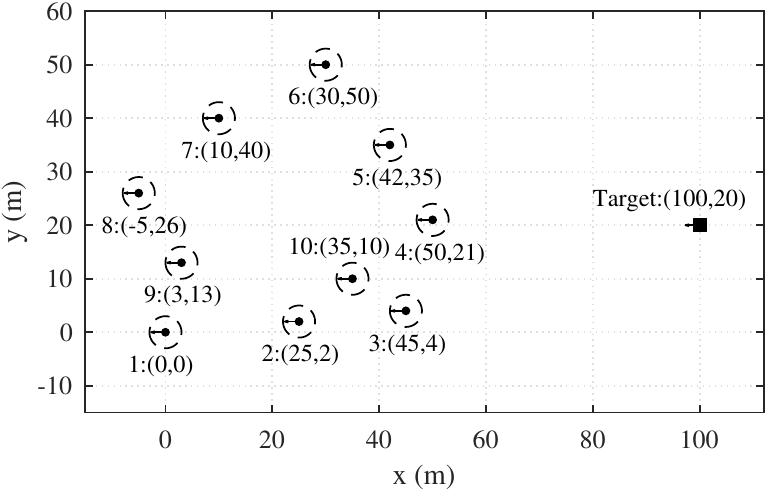} 
   \caption{The simulation scenario. The solid dots indicate the networked agents, the solid square denotes the target,
   and the arrows indicate the speed vector. The dotted circles denote the agent uncertainties. 
   The texts give the agent nominal positions in the form of agent index: $(x,y)$ m.
   }
   \label{fig:qin1}
\end{figure}

\begin{figure}
   \centering
   \subfloat[]{\includegraphics[width=0.48\linewidth]{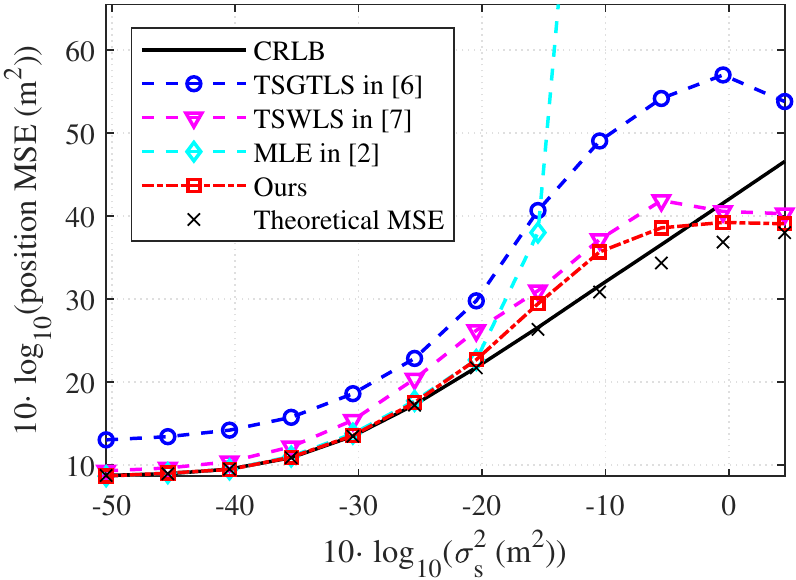}
   \label{fig:sim_pseudo-clock offset error}}
   \hfill
   \subfloat[]{\includegraphics[width=0.48\linewidth]{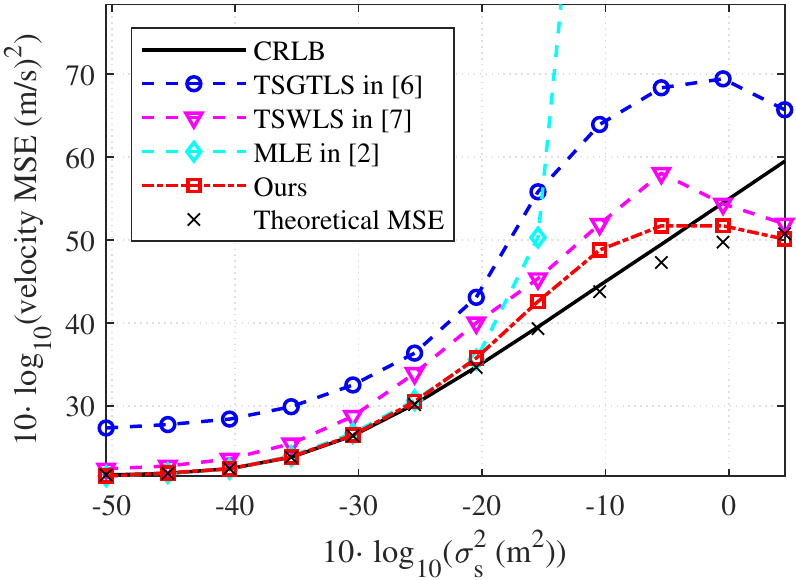}
   \label{fig:sim_clock skew error}}
   \caption{Performance comparison versus different agent uncertainties, under fixed topology of agents and target
   for non-LTCO scenarios.
   (a) Estimation results of moving target position.
   (b) Estimation results of moving target velocity.} 
   \label{fig:qin2}
\end{figure}

\textit{1) Non-LTCO scenarios:} 
The target clock offset is randomly generated from $-10$ to \mbox{$10$ ns},
to simulate the non-LTCO scenarios. 
\mbox{Fig. \ref{fig:qin2}} shows the estimation performance comparison versus agent uncertainties $\sigma_s^2$.
We can see that our proposed method significantly outperforms the existing methods.
Moreover, the performance of our method can attain the CRLB and consistent with theoretical MSE 
under small noises, which validates the performance analysis results.
TSGTLS and TSWLS perform worse mainly because they do not deeply exploit the nonlinear 
relationship between target parameters and nuisance parameters. 
Our method is also robust to large agent uncertainties, while MLE will diverge
due to large errors in its coefficient matrix $\mathbf A$.

\begin{figure} 
   \centering 
   \includegraphics[width=0.92\linewidth]{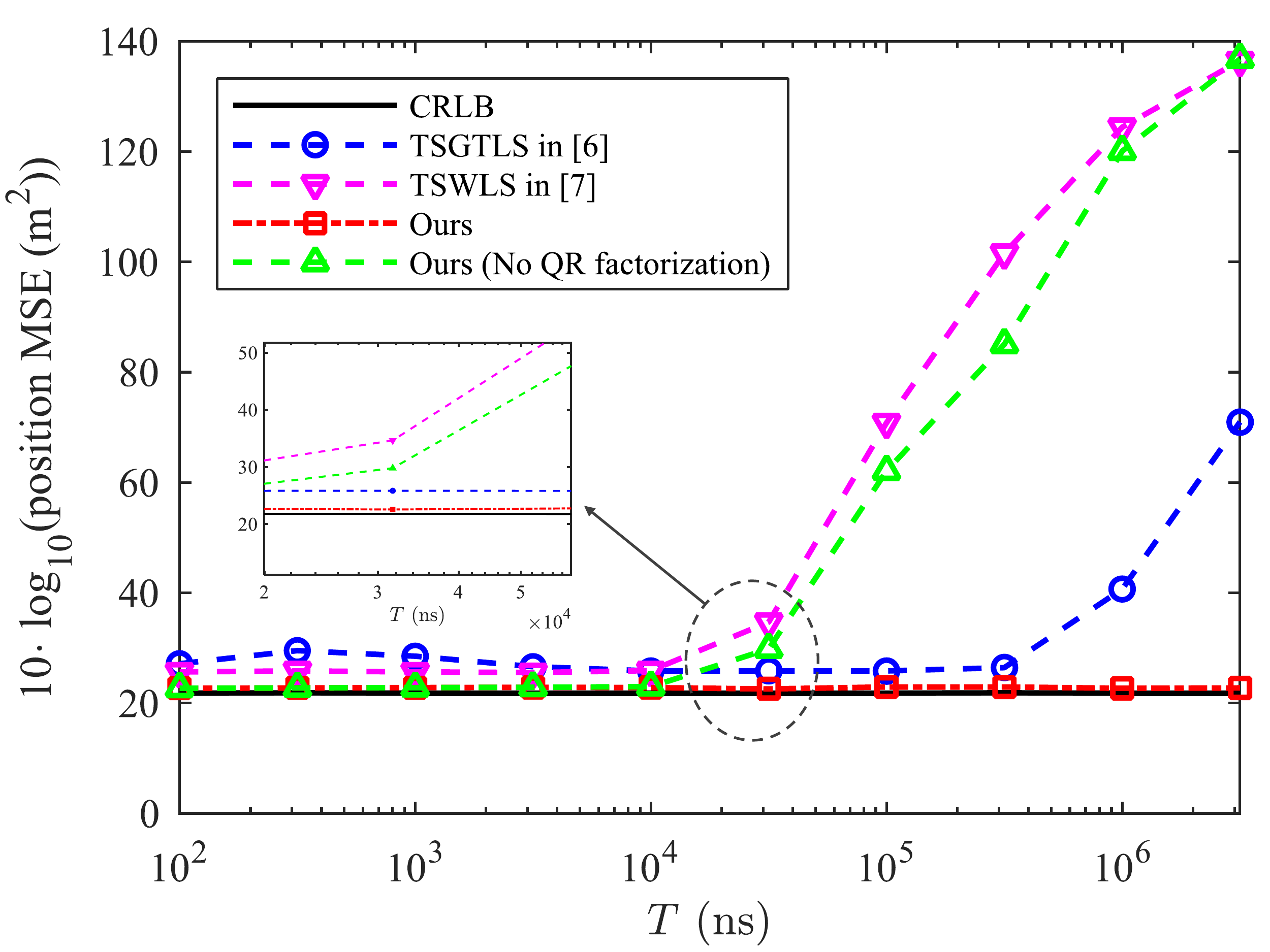} 
   \caption{Performance comparison versus different target clock offset, under fixed topology of agents and target.} 
   \label{fig:qin3}
\end{figure}

\begin{figure} 
   \centering 
   \includegraphics[width=0.89\linewidth]{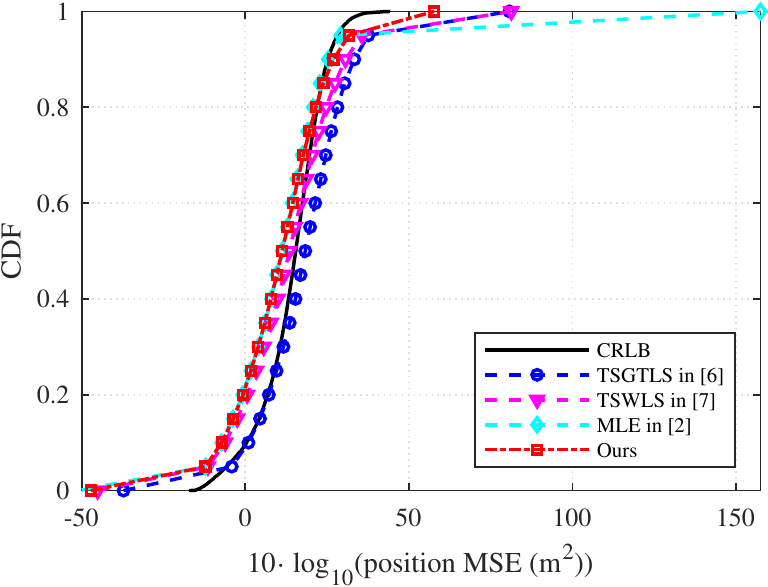} 
   \caption{Performance comparison under random topologies of agents and target.}
   \label{fig:qin4}
\end{figure}

\textit{2) LTCO scenarios:} 
We increase the target clock offset to simulate a weak to strong LTCO scenario
and fix \mbox{$\sigma_s = -20.5$ dB}. 
We note that when a target cannot access the multi-agent network, it may synchronize its clock via the 
Internet, in a typical precision of a few milliseconds \cite{mills1991internet}. 
When it enters the network, the strong LTCO scenario arises.
\mbox{Fig. \ref{fig:qin3}} demonstrates that our method is robust to strong LTCO scenarios
due to the advantage of utilizing the QR factorization, while the existing closed-form solutions fail.

\textit{3) Random topology:} 
For further evaluation, we randomly generate the agents and target topology
in each independent run, and fix \mbox{$\sigma_s = -20.5$ dB}, in non-LTCO scenarios.
The agent $x$ and $y$ coordinates are randomly drawn from $0$ to \mbox{$50$ m}, 
the target coordinates are from $-50$ to \mbox{$100$ m}, and their velocities in all directions are 
from $-5$ to \mbox{$5$ m/s}.
The number of independent runs is $N=10000$.
\mbox{Fig. \ref{fig:qin4}} shows the cumulative distribution function (CDF) of the position MSE.
We observe that the CDF curve of our method is steeper for most of the time and can achieve 1 under small position MSE,
which confirms that our method performs better than existing methods and is
more robust to random localization topology. 

The estimation results of clock offset, clock skew and/or target speed lead to the same conclusion, and
are not shown here due to space limitations.

\section{Conclusion}
In this letter, an extended TSWLS method is proposed to localize moving targets 
with the assistance of a multi-agent network. 
The position and velocity of the target are jointly estimated using the sequential TOA measurements.
Moreover, numerical stability is enhanced to produce valid estimates in LTCO scenarios.
Performance analysis and numerical results validate that our proposed method is approximately unbiased and reaches CRLB
under small noises.
Numerical simulations also show that our method outperforms the existing closed-form solutions in terms of accuracy and robustness,
both in non-LTCO and LTCO scenarios.

\bibliographystyle{IEEEtran}
\bibliography{IEEEabrv,ref}
\end{document}